\documentstyle[12pt]{article}
\begin{document}
\def\a{\alpha}\def\b{\beta}\def\g{\gamma}\def\d{\delta}\def\e{\epsilon}
\def\k{\kappa}\def\l{\lambda}\def\L{\Lambda}\def\s{\sigma}\def\S{\Sigma}
\def\Th{\Theta}\def\th{\theta}\def\om{\omega}\def\Om{\Omega}\def\G{\Gamma}
\def\y{\vartheta}\def\m{\mu}\def\n{\nu}
\def\ws{worldsheet}
\def\susy{supersymmetry}
\def\ts{target superspace}
\def\ks{$\k$--symmetry}
\renewcommand\baselinestretch{1.2}
\newcommand{\nn}{\nonumber\\}\newcommand{\p}[1]{(\ref{#1})}
\renewcommand{\thefootnote}{\arabic{footnote}}

\vspace*{3cm}
\begin{center}
{\large \bf  Duality-Symmetric Three-Brane and its Coupling
\\ to Type IIB Supergravity
}
\vspace{1cm}
\footnote{Work supported in part by the
INTAS Grants No 93-493-ext, No 96-0308
and by the Grant of Ukrainian State Committee on Science and Technology
.  }\\
Alexei Nurmagambetov \footnote{e-mail:
ajn@kipt.kharkov.ua}\\
\vspace{0.5cm}
{\it Kharkov
Institute of Physics and Technology}\\
{\it 310108, Kharkov, Ukraine}\\
\bigskip

\vspace{1.5cm}
{\bf Abstract}
\end{center}

Starting from the bosonic sector of the M-theory super-five-brane we
obtain the action for duality-symmetric three-brane and
construct the consistent coupling of
the proposed action with the bosonic sector of type IIB supergravity.

\vspace{0.8cm}

\vspace{0.8cm}

\newpage

\section{Introduction}

S-duality is conjectured to be an exact symmetry of the type IIB
superstring theory, whose BPS spectrum includes in particular
the IIB three-brane, being invariant under the action of S-duality
symmetry group.

The special role of the three-brane in the type IIB superstring theory has
been noted in literature (see, for example, \cite{duff} -- \cite{ryang})
time and again in connection with conjecture about possible reformulation
of the $D=10$ IIB superstring theory as a theory in twelve dimensions
\cite{hull}, \cite{ftheory}. In frames of this conjecture
the type IIB superstring theory
could be interpreted as a theory of fundamental supersymmetric
three-branes \cite{tseytlin}, \cite{jatkar}, whose $SL(2,Z)$ symmetry
would be a consequence of the `electro-magnetic` worldvolume duality, with
S-duality invariant effective action in type IIB theory background.
However, one of the obstacles in construction of this effective action is
the field content of type IIB supergravity \cite{2b}, whose bosonic sector
includes a self-dual antisymmetric gauge field.

The problem of a compatibility of the Lagrangian description of self-dual
fields and space-time covariance has attracted great deal of attention
during a long period of time \cite{sd}. The progress in construction
of the Lagrangians for self-dual antisymmetric gauge fields was achieved
by Schwarz and Sen \cite{ss}. Their results was the generalization of
early work by Floreanini and Jackiw \cite{fj} on $d=2$ chiral scalars and
by Henneaux and Teitelboim \cite{ht} on $d=2(p+1)$ chiral p-forms.
However, the approach of \cite{ss}, being manifestly
duality invariant, is not manifestly space-time covariant.
But it is desirable to have a covariant formulation, especially when one
is interested in consideration of models coupled to gravity or
supergravity.  This requires the use of auxiliary fields, from one to
infinity in dependence on the approach \cite{siegel}, \cite{infinity},
\cite{pst},\cite{pst1}. In what follows we will be concentrated on the
approach proposed by Pasti, Sorokin and Tonin (PST) \cite{pst},
\cite{pst1} with single auxiliary field ensuring the covariance of the
model, but entering to the action in a non-polynomial way. Recently this
approach was successfully applied to a different kind of models with
self-dual or duality-symmetric antisymmetric gauge fields including the
M-Theory super-five-brane \cite{blnpst}, \cite{sbr}, the bosonic sector of
$D=10$ type IIB supergravity \cite{als} and duality-symmetric $D=11$
supergravity \cite{bbs}. In this paper we apply the PST approach to obtain
the effective action of the type IIB superstring theory with three-brane
source.

As a preliminary step in Section 2 we obtain the action for
three-brane with duality-symmetric worldvolume gauge fields by
worldvolume dimensional reduction of the action for bosonic five-brane
\cite{5br}. This special structure of the three-brane action is dictated
by the structure of the bosonic sector of IIB supergravity \cite{als},
which contains a self-dual forth rank antisymmetric gauge field naturally
coupled to three-brane, having the special symmetries being characteristic
of theories with self-dual gauge fields \cite{pst}, \cite{pst1}. This
symmetries have to be preserved under the coupling of three-brane with the
type IIB supergravity that automatically demands the duality-symmetric
structure of three-brane as we will see later on. In Section 3 we review
the PST approach to the bosonic sector of $D=10$ IIB supergravity
\cite{als}, entering to the effective action of the type IIB superstrings,
and obtain the action describing consistent coupling of three-brane with
the bosonic sector of $D=10$ IIB supergravity. Discussion of the
obtained results and some concluding remarks are
collected in the last section.

\section{Three-brane from five-brane}

Our starting point is
the action for bosonic five-brane \cite{5br}:
\begin{equation}\label{1}
S=-\int\,d^6\xi\,\sqrt{-\det(g_{\mu\nu}+i{\tilde{H}}_{\mu\nu})}
+\sqrt{-g}{1\over 4{\sqrt{-{(\partial
a)}^2}}}{\tilde{H}}^{\mu\nu}H_{\mu\nu\rho} \partial^{\rho}a \end{equation}
with
\begin{equation}\label{2}
{\tilde{H}}_{\mu\nu}\equiv {1\over \sqrt{-{(\partial
a)}^2}}H^{*}_{\mu\nu\rho}\partial^{\rho} a;\ \ \ H^{*\mu\nu\l}={1\over
{3!\sqrt{-g}}}\epsilon^{\m\n\l\rho\s\d}H_{\rho\s\d},
\end{equation}
where
$H_{\mu\nu\l}$ is the field strength of the worldvolume antisymmetric
tensor field $B_{\m\n}(\xi)$ and $g_{\m\n}$ is the metric in six
dimensions.

To obtain the action describing duality-symmetric three-brane we
perform a dimensional reduction by splitting the
six-dimensional indices $\m$ onto the set of $(m,\a)$ \cite{mahsch}, where
$m$ is the four-dimensional index and $\a$ is the index in compact
dimensions. We will assume $a\not= a(y^\a)$, where
$y^\a$ are the coordinates of a compact space; $B_{mn}=0$,
$B_{\rho p}\not=B_{\rho p}(y^\a)$ and $B_{\a\b}=0$;
and will choose the standard
form of vielbeins \cite{scherksch}, \cite{mahsch} that allows us to write
the six-dimensional Levi-Civita symbol $\epsilon_{\m\n\rho\s\d}$ as the
direct product of the four-dimensional Levi-Civita symbol $\e_{mnrq}$ and
the unit antisymmetric tensor $\e_{\a\b}$ in compact dimensions.

Using this ansatz, after some algebra we obtain the
following action for three-brane with duality-symmetric worldvolume gauge
fields:
$$
S=-{\cal{V}}\int\,d^4\xi\,\sqrt{-G}(\sqrt{1+{\tilde{F}}_{m\a}
{\tilde{F}}^{m\a}
+{1\over 4}
(\e^{abdf}{\tilde{F}}_{b\a}\e^{\a\b}{\tilde{F}}_{d\b}{{\partial_f a}\over
\sqrt{-{(\partial a)}^2}})^2}
$$
\begin{equation}\label{3}
-{1\over 2}{\tilde{F}}_{m\a}\e^{\a\b}F^{mn}_{\b}
{\partial_n a \over{\sqrt{
-(\partial a)^2}}}),
\end{equation}
where
\begin{equation}\label{4}
{\tilde{F}}_{m\a}={{\partial^n a}\over {\sqrt{-{(\partial a)}^2}}}
F^*_{nm\a};\ \ \ F^{\b}_{mn}=2\partial_{[m}B^{\b}_{n]};
\ \ \
\sqrt{-G}=\sqrt{-\det G_{mn}},
\end{equation}
the metric tensor for the internal space is $g_{\a\b}=\d_{\a\b}$,
and ${\cal{V}}$ is the volume of a compact space.

The action \p{3}, written in manifestly covariant form, is invariant
under
\begin{itemize}
\item
worldvolume diffeomorphisms,
\item
usual gauge invariance
\begin{equation}\label{5}
\d B^{\a}_{m}=\partial_{m}\phi^{\a},
\end{equation}
\item
local transformations of the form
\begin{equation}\label{6}
\d B^{\a}_{m}=\partial_{m}a(\xi)\varphi^{\a}(\xi);\ \ \ \d a=0,
\end{equation}
\item
and additional local symmetry
\begin{equation}\label{7}
\d a(\xi)=\Phi(\xi),\ \ \   \d B^{\a}_{m}={\Phi(\xi)\over
(\partial a)^2}\e^{\a\b}{\cal{F}}_{m\b},\ \ \ {\cal{F}}_{m}^\a=
{\cal{V}}^{\a}_{m}-\e^{\a\b}F_{mn\b}\partial^n a,
\end{equation}
\end{itemize}
where
\begin{equation}\label{8}
{\cal{V}}^{\a}_{m}=\sqrt{-{(\partial a)}^2}\ \ {\d \ \
\sqrt{1+{\tilde{F}}_{m\a}{\tilde{F}}^{m\a}+{1\over 4}
(\e^{abdf}{\tilde{F}}_{b\a}\e^{\a\b}{\tilde{F}}_{d\b}{{\partial_f a}\over
\sqrt{-{(\partial a)}^2}})^2}
\over \d {{\tilde{F}}_{\a}}^{m}}.
\end{equation}
All these symmetries become obvious if we consider a variation of the
action \p{3} under $B^{\a}_m$ and $a(\xi)$, written in differential forms
as
$$
\d S=-{\cal{V}}\int_{{\cal{M}}^4}\, \, (\d B^{(1)\a}-{\d a \over
{(\partial a)}^2}\e^{\a\b}{\cal{F}}^{(1)}_\b)\wedge
d({1\over\sqrt{-{(\partial a)}^2}} da\wedge {\cal{F}}^{(1)}_\a).
$$

The presence of the symmetry \p{7} is crucial for establishing a connection
with non-covariant formalism
of \cite{berman}, \cite{parra}. Effectively, the symmetry \p{7} allows one to
fix a gauge in such a way
that, say,
\begin{equation}\label{9}
\partial_{m} a(\xi)=\d^0_m,
\end{equation}
or
\begin{equation}\label{10}
\partial_{m} a(\xi)=\d^3_m,
\end{equation}
and the local symmetry \p{7} ensures the auxiliary role of the field 
$a(\xi)$.  Thus, fixing the gauge \p{9} or \p{10}, we can eliminate 
$a(\xi)$ from the action without losing dynamical information and obtain 
the action \footnote{The choice \p{9} corresponds to the action (15) of 
\cite{parra} and as it has noted in \cite{berman}, the choice \p{10} leads 
to the action with a pair of electric fields.} similar to that of 
\cite{berman} and \cite{parra}.

The symmetry \p{6} allows one to reduce the general solution for the
equations of motion of $B^{\a}_{m}$ fields to the form
\begin{equation}\label{11}
{\cal{V}}^\a_m-\e^{\a\b}F_{mn\b}\partial^n a=0.
\end{equation}
Eq. \p{11} is a generalization of the self -- duality condition to the
case of self-interacting gauge fields \cite{perry}. In the linearized
limit, where we take the flat metric and insert in \p{3} the term
$-{1\over 2}\tilde{F}_{m \a}\tilde{F}^{m \a}$ instead of the square root,
the condition \p{11} becomes the self-duality condition for
duality-symmetric Maxwell electrodynamics \cite{pst}
$$
{\cal{F}}^{\a}_{mn}=0
$$
with
$$
{\cal{F}}^{\a}_{mn}=\e^{\a\b}F_{\b mn}-F_{mn}^{*\a}.
$$

Thus, the action \p{3} describes a duality-symmetric three-brane.
It involves a pair of a gauge fields $B^\a_m$
related to each other by virtue of the self-duality condition \p{11}.
After elimination of one of the Abelian fields we recover the usual
Born-Infeld action as it was shown in \cite{berman}, \cite{parra}. The
presence of an additional field $a(\xi)$ allows one to covariantize the
approach of \cite{berman}, \cite{parra} and to make electric-magnetic
duality manifest at the level of action.

To obtain the three-brane source term of the effective action for type IIB
theory we have to consider three-brane coupling with
IIB $D=10$ background fields, including the dilaton
$\phi$, the second rank antisymmetric tensor
$C_{\underline{m}\underline{n}}$ and the metric
$g_{\underline{m}\underline{n}}$ in NS-NS sector and the axion $\chi$, the
second rank antisymmetric tensor $\tilde{C}_{\underline{m}\underline{n}}$
together with the self-dual fourth rank antisymmetric tensor
$A_{\underline{m}\underline{n}\underline{p}\underline{q}}$ in RR sector
(${\underline{m}}, {\underline{n}}, {\underline{p}},
{\underline{q}}=0,\dots,9$).

First consider the action \p{3} in dilaton-axion background, describing
by the $2\times 2$ matrix valued
scalar field
\begin{equation}\label{13}
{\cal{M}}={1\over \l_2(x)}
\left(\begin{array}{cc} 1 & \l_1(x)\\
\l_1(x) & {\l_1}^2+{\l_2}^2
\end{array}\right),
\end{equation}
\begin{equation}\label{14}
\l=\l_1+i\l_2\equiv \chi+ie^{-\phi},\ \ \phi=\phi(x(\xi)),\ \
\chi=\chi(x(\xi)),
\end{equation}
satisfying the following conditions:
\begin{equation}\label{15}
{\cal{M}}^T={\cal{M}};\ \ \ {\cal{M}}\e {\cal{M}}^T=\e.
\end{equation}

${\cal{M}}$, $\e$ and $B^\a_m$ transform under the global $SL(2,R)$
transformations $\omega$ as follows:
\begin{equation}\label{16}
{\cal{M}}\longrightarrow \omega^T {\cal{M}}\omega,
\ \ \omega \e \omega^T=\e,\ \ B^\a_m=(\omega^T B)^\a_m,
\end{equation}
$$
\omega=
\left(\begin{array}{cc} A & B\\
C & D
\end{array}\right),\ \ \ AD-BC=1.
$$
Then the $SL(2,R)$ invariant action has the following form
\footnote{To check this invariance, the transformation law
${\tilde{F}}^\a_m=({\tilde{F}}\omega)^\a_m$ has to be used.}:
\begin{equation}\label{17}
S=-{\cal{V}}\,\int\,d^4\xi\,\sqrt{-G}(\sqrt{1+{\tilde{F}}_m^\a
(\e^T{\cal{M}}\e)_{\a\b}
{\tilde{F}}^{\b m}+{1\over 4}
(\e^{abdf}{\tilde{F}}_{b\a}\e^{\a\b}{\tilde{F}}_{d\b}{{\partial_f a}\over
\sqrt{-{(\partial a)}^2}})^2}
\end{equation}
$$
-{1\over 2}{{\tilde{F}}^m}_\a\e^{\a\b}F_{mn\b}{\partial^n a \over{\sqrt{
-(\partial a)^2}}}).
$$

To find the action \p{17} in a
background of antisymmetric gauge fields of IIB $D=10$
supergravity, we have to replace the field
strength $F^\a_{mn}$ with
\begin{equation}\label{18}
H^\a_{mn}=F^\a_{mn}-C^\a_{mn}
\end{equation}
and add to the action \p{17} a Wess-Zumino term (see \cite{tseytlin},
\cite{schwarz}, \cite{5br}). The resulting action becomes
$$
S=-{\cal{V}}\int\,d^4\xi\,\sqrt{-G}(\sqrt{1+{\tilde{H}}_m^\a
(\e^T{\cal{M}}\e)_{\a\b}
{\tilde{H}}^{\b m}+{1\over 4}
(\e^{abdf}{\tilde{H}}_{b\a}\e^{\a\b}{\tilde{H}}_{d\b}{{\partial_f a}\over
\sqrt{-{(\partial a)}^2}})^2}
$$
\begin{equation}\label{19}
-{1\over 2}{{\tilde{H}}^m}_{\a}\e^{\a\b}H_{mn\b}{\partial^n a \over{\sqrt{
-(\partial a)^2}}})
+\int_{{\cal{M}}^4}\,(A^{(4)}+{1\over 2}\e^{\a\b}F_\a^{(2)}\wedge
C_\b^{(2)}),
\end{equation}
where $A^{(4)}$ and $C^{(2)}$ are the pullbacks of the corresponding D=10
forms onto worldvolume ${\cal{M}}^4$.

The structure of the Wess -- Zumino term is governed by the requirement of
invariance of \p{19} under the modified
symmetries \p{5} -- \p{7}
$$
\d a=0;\ \ \ \d B^{\a}_m=\partial_m\phi^{\a}+
\partial_m a\varphi^{\a},
$$
\begin{equation}\label{12}
\d a(\xi)=\Phi(\xi),\ \ \   \d B^{\a}_{m}={\Phi(\xi)\over
(\partial a)^2}\e^{\a\b}{\cal{H}}_{m\b},\ \ \ {\cal{H}}_{m}^\a=
\tilde{\cal{V}}^{\a}_{m}-\e^{\a\gamma}H_{mn\gamma}\partial^n a
\end{equation}
with
$$
\tilde{\cal{V}}^{\a}_{m}=\sqrt{-{(\partial a)}^2}\ \ {\d \ \
\sqrt{1+{\tilde{H}}_m^\a
(\e^T{\cal{M}}\e)_{\a\b}
{\tilde{H}}^{\b m}+{1\over 4}
(\e^{abdf}{\tilde{H}}_{b\a}\e^{\a\b}{\tilde{H}}_{d\b}{{\partial_f a}\over
\sqrt{-{(\partial a)}^2}})^2}
\over \d {{\tilde{H}}_{\a}}^{m}},
$$
that is, the Wess --
Zumino term is required to preserve local symmetries of the action when
three -- brane couples to the antisymmetric fields (see \cite{5br}, where
this fact was pointed out for the first time).

To complete a description, let us note that the action \p{19} is $SL(2,R)$
invariant under the following $SL(2,R)$ transformation of the background
fields:
\begin{equation}\label{20}
C^\a_{mn}=
\left(\begin{array}{c} C_{mn}\\ \tilde{C}_{mn}
\end{array}\right)\longrightarrow
(\omega^T C)^\a_{mn}.
\end{equation}

\section{D=10 IIB supergravity and its coupling to three-brane}

Although the Lorentz covariant equations of motion for $D=10$ IIB
supergravity are well known for a long time \cite{2b}, the problem of
construction of the complete $D=10$ IIB supergravity action functional
\cite{kavalov} remains open. However, from the modern point of view
the main obstacle in construction of this action connected with the
presence of a self-dual antisymmetric gauge field in the bosonic sector of
Type IIB supergravity has been avoided in the work by Dall'Agata, Lechner
and Sorokin \cite{als} where the PST approach \cite{pst,pst1} was applied.
In this section we review the approach of \cite{als} and
obtain the effective action for the Type IIB superstring theory with the
three-brane source.

The bosonic sector of the type IIB supergravity \cite{2b} is described by
the following action \cite{als}:
$$
S=\int\, d^{10}x \sqrt{-g}[R-2\partial_{\underline{m}}\phi
\partial^{\underline{m}}\phi-2e^{2\phi}\partial_{\underline{m}}\chi
\partial^{\underline{m}}\chi-{1\over 3}e^{-\phi}H_{\underline{lmn}}
H^{\underline{lmn}}
$$
$$
-{1\over 3}e^{\phi}(\tilde{H}_{\underline{lmn}}
-\chi H_{\underline{lmn}})(\tilde{H}^{\underline{lmn}}
-\chi H^{\underline{lmn}})]
$$
\begin{equation}\label{101}
-{1\over 6}\int\, d^{10}x {\sqrt{-g}\over {\partial_{\underline{r}}a
\partial^{\underline{r}}a}}\partial^{\underline{l}}a(x)
M^{*}_{\underline{l}\underline{m_1}\dots\underline{m_4}}
{\cal{M}}^{\underline{m_1}\dots\underline{m_4}\underline{p}}
\partial_{\underline{p}}a(x)
-4\int_{{\cal{M}}^{10}}\, A^{(4)}\wedge H \wedge \tilde{H},
\end{equation}
where
$R(x)$ is a $D=10$ scalar curvature,
$H^{(2)}=d C^{(2)}$ and $\tilde{H}^{(2)}=d \tilde{C}^{(2)}$ are,
respectively, the field strength of the NS-NS and R-R two-forms;
\begin{equation}\label{102}
M^{(5)}=d A^{(4)}+{1\over 2}C^{(2)}\wedge d \tilde{C}^{(2)}
-{1\over 2}\tilde{C}^{(2)}\wedge d C^{(2)}
\end{equation}
is the five-form field strength of the R-R self-dual gauge field
$A^{(4)}(x)$, ${\cal{M}}^{(5)}=M^{(5)}-M^{*(5)}$ is the anti-self-dual
part of $M^{(5)}(x)$, and the auxiliary scalar field $a(x)$ ensures the
manifest $D=10$ covariance of the $A^{(4)}$ part of the action.

The action \p{101} possesses the following symmetries:
\begin{itemize}
\item
$D=10$ general covariance;
\item
usual gauge invariance
$$
\d C^{(2)}=d \a^{(1)},\ \ \ \d \tilde{C}^{(2)}=d \tilde{\a}^{(1)},
$$
\begin{equation}\label{103}
\d A^{(4)}=d\phi^{(3)}-{1\over 2}d \a^{(1)}\wedge \tilde{C}^{(2)}
+{1\over 2}d \tilde{\a}^{(1)}\wedge C^{(2)};
\end{equation}
\item
global $SL(2,R)$ symmetry mixing of $\phi$ and $\chi$ and of $C^{(2)}$ and
$\tilde{C}^{(2)}$
\item
and additional local symmetries:
$$
\d A^{(4)}=da\wedge \varphi^{(3)},\ \ \ \d a=0,
$$
\begin{equation}\label{104}
\d a=\Phi(x),\ \ \ \d A^{(4)}_{\underline{mnpq}}={\d a\over {(\partial
a)^2}}{\cal{M}}^{(5)}_{\underline{mnpqr}}\partial^{\underline{r}}a.
\end{equation}
\end{itemize}

Apparently, the kinetic terms involving $H^{(3)}$ and $\tilde{H}^{(3)}$
are the gauge invariant. The invariance under the symmetries corresponding
to the transformations of $A^{(4)}$ \p{103} and \p{104} can be established
by the variation of the last two terms of the action \p{101} having the
following form in the differential forms notations:
$$
\d S_{M+CS}=-4\int_{{\cal{M}}^{10}}\,\, \d M^{(5)}\wedge M^{(5)}+
2\d M^{(5)}\wedge v \wedge i_{v}{\cal{M}}^{(5)}+\d S_{CS}+\d_{a(x)}S_M
$$
$$
=-4\int_{{\cal{M}}^{10}}\,\,[2({\d a\over{\sqrt{-(\partial a)^2}}}
i_{v}{\cal{M}}^{(5)}+\d A^{(4)}+{1\over 2}C^{(2)}\wedge \d \tilde{C}^{(2)}
-{1\over 2}\tilde{C}^{(2)}\wedge \d C^{(2)})
\wedge d(v\wedge i_{v}{\cal{M}}^{(5)})
$$
$$
+2(d\tilde{C}^{(2)}\wedge \d C^{(2)}-dC^{(2)}\wedge \d\tilde{C}^{(2)})
\wedge(v\wedge i_{v}{\cal{M}}^{(5)}-{1\over 2}M^{(5)})
$$
$$
+dM^{(5)}\wedge(\d A^{(4)}+{1\over 2}C^{(2)}\wedge \d \tilde{C}^{(2)}
-{1\over 2}\tilde{C}^{(2)}\wedge \d C^{(2)})
$$
\begin{equation}\label{105}
+\d A^{(4)}\wedge dC^{(2)}\wedge d\tilde{C}^{(2)}-
\d A^{(4)}\wedge d\tilde{C}^{(2)}\wedge \d C^{(2)}
+d A^{(4)}\wedge dC^{(2)}\wedge \d\tilde{C}^{(2)}],
\end{equation}
where $v={da\over{\sqrt{-(\partial a)^2}}}$ and the contraction
$i_{v}{\cal{M}}^{(5)}$ is defined as
\begin{equation}\label{106}
i_{v}{\cal{M}}^{(5)}={1\over 4!}dx^{\underline{m_1}}\wedge\dots\wedge
dx^{\underline{m_4}}v^{\underline{p}}{\cal{M}}_{\underline{p m_4}\dots
\underline{m_1}}.
\end{equation}

The first line of the \p{105} makes the gauge invariance obvious since
$\d_{gauge} M^{(5)}=0$, and $\d_{gauge} S_{CS}=0$ up to the total
derivative. Concerning the invariance under the transformations \p{104},
it should be noted that the form of the Chern-Simons term is completely
determined by the requirement of keeping the symmetries \p{104} and by the
form of $M^{(5)}$. On the other hand, the form of the Chern-Simons term is
fixed by the requirement of supersymmetry. Thus, the symmetry \p{104} is
the criterion of selfconsistency for the bosonic theories with the
self-dual fields playing the same role as supersymmetry in superfield
theories or as kappa-symmetry for super-p-branes.

As for the global $SL(2,R)$, the presence of this symmetry becomes clear
if we write down the action \p{101} using the matrix \p{13} with
$\phi=\phi(x)$ and $\chi=\chi(x)$:
$$
S=\int\, d^{10}x \sqrt{-g}[R-g_{\underline{mn}}tr
(\partial^{\underline{m}}{\cal{M}}\e
\partial^{\underline{n}}{\cal{M}}\e)-{1\over 3}
{\cal{H}}^\a_{\underline{lmn}}(\e^T{\cal{M}}\e)_{\a\b}
{\cal{H}}^{\b \underline{lmn}}
$$
\begin{equation}\label{107}
-{1\over 6}\int\, d^{10}x {\sqrt{-g}\over {\partial_{\underline{r}}a
\partial^{\underline{r}}a}}\partial^{\underline{l}}a(x)
M^{*}_{\underline{l}\underline{m_1}\dots\underline{m_4}}
{\cal{M}}^{\underline{m_1}\dots\underline{m_4}\underline{p}}
\partial_{\underline{p}}a(x)
+2\int_{{\cal{M}}^{10}}\, A^{(4)}\wedge {\cal{H}}^\a \e_{\a\b} \wedge
{\cal{H}}^\b,
\end{equation}
where
\begin{equation}\label{1071}
{\cal{H}}^\a=
\left(\begin{array}{c} dC\\ d\tilde{C}
\end{array}\right),\ \
\e_{\a\b}=
\left(\begin{array}{cc} 0 & -1\\
1 & 0
\end{array}\right),
\end{equation}
and taking into account the transformation laws for ${\cal{M}}$ \p{15} and
for $C^{(2)}$ \p{20}.

Equation of motion for $A^{(4)}$ field is reduced by \p{104} to the
self-duality condition
\begin{equation}\label{108}
{\cal{M}}^{(5)}=M^{(5)}-M^{*(5)}=0,
\end{equation}
provided a fulfillment of the equation of motion
\begin{equation}\label{109}
d M^{*(5)}=-dC^{(2)}\wedge d\tilde{C}^{(2)},
\end{equation}
derived early in \cite{2b}.

To construct a consistent coupling of three-brane to type IIB supergravity
let us firstly analyze the equations \p{108} and \p{109} following from
the action \p{101}. The self-duality of $M^{(5)}$ provides the symmetry
between equation of motion for $M^{(5)}$ and its Bianchi identities:
\begin{equation}\label{111}
d M^{*(5)}=-dC^{(2)}\wedge d\tilde{C}^{(2)},\ \ \
dM^{(5)}=-dC^{(2)}\wedge d\tilde{C}^{(2)},
\end{equation}
hence, if one would like to introduce sources into these equations
\cite{berkovits}, one should follow the prescription used by Dirac for
describing magnetic monopoles coupled to D=4 Maxwell fields. Following
Dirac we can provide a fulfillment of the Bianchi identities
\begin{equation}\label{112}
d\hat{M}=0,
\end{equation}
replacing $M^{(5)}$ with $\hat{M}^{(5)}=M^{(5)}-G^{*(5)}$, where
\begin{equation}\label{113}
dG^{*(5)}={1\over 4}J^{*(4)},
\end{equation}
and consider the generalization of the Dirac
string to a Dirac four-brane stemmed from the three-brane.

Following \cite{bbs} we extend the action for three-brane to an integral
over $D=10$ space-time by inserting a $\d$-function closed six-form
\begin{equation}\label{114}
J^{*(4)}={1\over{4!6!}}dx^{\underline{m_1}}\wedge\dots\wedge
dx^{\underline{m_6}}\e_{\underline{m_1}\dots\underline{m_6}\underline{n_1}
\dots\underline{n_4}}\int_{{\cal{M}}^4}\,d \hat{x}^{\underline{n_1}}\wedge
\dots\wedge d \hat{x}^{\underline{n_4}} \d(x-\hat{x}(\xi))
\end{equation}
dual to the current $J^{(4)}$ minimally coupled to $A^{(4)}(x)$. Then the
action describing consistent coupling is
$$
S=\int\, d^{10}x \sqrt{-g}[R-g_{\underline{mn}}tr
(\partial^{\underline{m}}{\cal{M}}\e
\partial^{\underline{n}}{\cal{M}}\e)-{1\over 3}
{\cal{H}}^\a_{\underline{lmn}}(\e^T{\cal{M}}\e)_{\a\b}
{\cal{H}}^{\b \underline{lmn}}
$$
$$
-{1\over 6}\int\, d^{10}x {\sqrt{-g}\over {\partial_{\underline{r}}a
\partial^{\underline{r}}a}}\partial^{\underline{l}}a(x)
\hat{M}^{*}_{\underline{l}\underline{m_1}\dots\underline{m_4}}
\hat{{\cal{M}}}^{\underline{m_1}\dots\underline{m_4}\underline{p}}
\partial_{\underline{p}}a(x)
+2\int_{{\cal{M}}^{10}}\, A^{(4)}\wedge {\cal{H}}^{(2)\a} \e_{\a\b} \wedge
{\cal{H}}^{(2)\b}
$$
$$
-{\cal{V}}\int\,d^4\xi\,\sqrt{-G}(\sqrt{1+{\tilde{H}}_m^\a
(\e^T{\cal{M}}\e)_{\a\b}
{\tilde{H}}^{\b m}+{1\over 4}
(\e^{abdf}{\tilde{H}}_{b\a}\e^{\a\b}{\tilde{H}}_{d\b}{{\partial_f a}\over
\sqrt{-{(\partial a)}^2}})^2}
$$
\begin{equation}\label{115}
-{1\over 2}{{\tilde{H}}^m}_{\a}\e^{\a\b}H_{mn\b}{\partial^n a \over{\sqrt{
-(\partial a)^2}}})
+\int_{{\cal{M}}^4}\, A^{(4)}
-2\int_{{\cal{M}}^{10}}\,
\e^{\a\b}H_\a^{(2)}\wedge
dC_\b^{(2)}\wedge G^{*(5)},
\end{equation}
where $\hat{M}^{(5)}=M^{(5)}-G^{*(5)}$ and
$\hat{{\cal{M}}}^{(5)}=\hat{M}^{(5)}-\hat{M}^{*(5)}$.

The requirement of consistency demands preservation of the local
symmetries \p{12} and \p{104} as well as global $SL(2,R)$. This puts the
restriction to the additional scalar field $a(x)$ to be
the image onto the worldvolume of three-brane, i.e.
$a=a(x(\xi))$ (see \cite{bbs} for details).

Thus, we have constructed the effective action for type IIB supergravity
with the three-brane source.

\section{Discussion}

To summarize, we start from the action for the bosonic sector of the
M-theory super-five-brane and obtain the action for three-brane with
duality-symmetric worldvolume gauge fields. This fields are related to
each other by virtue of the self-duality condition and elimination of one
of them yields the usual Born-Infeld action.
In the presence of the IIB $D=10$ background fields the proposed action
possesses an $SL(2,R)$ symmetry and the requirement of keeping the
additional local symmetries characteristic of the PST approach to the
theories with chiral gauge fields completely restricts the structure of
the Wess-Zumino term.

The duality-symmetric structure of the three-brane action allows to
construct the consistent coupling to the type IIB supergravity preserving
all of the local and global symmetries of the effective action. It leads
to appearance of minimal term coupled to $A^{(4)}$ as well as to the new
types of non-minimal terms such as
$\e^{\a\b}H_\a^{(2)}\wedge dC_\b^{(2)}\wedge G^{*(5)}$ that reflects the
presence of electric and magnetic charges carrying by the worldvolume of
three-brane.

The straightforward developments of the obtained results are the
investigation of the equations of motion of the effective action,
the anomaly analysis \cite{yin} and the construction of
the supersymmetric generalization of the proposed action for three-brane
in a spirit of \cite{berman1}.

The investigation of these and other questions we postpone to future
papers.

\vspace{1cm}
{\underline{\bf Acknowledgements.}} The author is grateful to Dmitri
Sorokin and to Igor Bandos for interest to this work and
useful remarks and suggestions.
I also wish to acknowledge the
hospitality of the International Centre for Theoretical Physics
where the main part of this work was complete.

\end{document}